\documentclass[aps,superscriptaddress,prl,fleqn,showpacs,nofootinbib,preprintnumbers]{revtex4}
\usepackage{amssymb,amsmath,epsfig}

\begin{document}

\title{Generalized TMDs and the exclusive double Drell-Yan process}

\author{Shohini Bhattacharya}
\affiliation{Department of Physics, SERC,
             Temple University, Philadelphia, PA 19122, USA}

\author{Andreas Metz}
\affiliation{Department of Physics, SERC,
             Temple University, Philadelphia, PA 19122, USA}
             
\author{Jian Zhou}
\affiliation{School of Physics and Key Laboratory of Particle Physics and Particle Irradiation (MOE),
              Shandong University, Jinan, Shandong 250100, China}

\begin{abstract}
Generalized transverse momentum dependent parton distributions (GTMDs) are the most general parton correlation functions of hadrons.
By considering the exclusive double Drell-Yan process it is shown for the first time how quark GTMDs can be measured.
Specific GTMDs can be addressed by means of polarization observables.
\end{abstract}


\date{\today}

\maketitle

\section{Introduction} 
\label{s:intro}
Multi-dimensional imaging of strongly interacting systems is currently a very active research area. 
The key quantities of this field are new types of parton distribution functions (PDFs) which are extensions of the one-dimensional PDFs that became textbook material --- generalized parton distributions (GPDs) and transverse momentum dependent parton distributions (TMDs), which provide multi-dimensional images of hadrons in position space and momentum space, respectively.
Studying GPDs and TMDs is a core mission at several particle accelerator facilities worldwide and, in particular, at a potential future electron-ion collider (EIC)~\cite{Boer:2011fh,Accardi:2012qut}.

In this context, GTMDs~\cite{Meissner:2008ay,Meissner:2009ww,Lorce:2013pza} have recently attracted enormous interest.
Since several GTMDs reduce to GPDs and TMDs in certain kinematical limits, they are often denoted as ``mother distributions."
The Fourier transform of GTMDs are Wigner functions~\cite{Ji:2003ak,Belitsky:2003nz}, the quantum-mechanical counterpart of classical phase-space distributions.
Partonic Wigner functions contain information on the five-dimensional parton structure --- the (average) longitudinal and transverse momentum as well as transverse position of partons inside a hadron~\cite{Lorce:2011dv}.
Two of the GTMDs --- $F_{1,4}$ and $G_{1,1}$ in the notation of~\cite{Meissner:2009ww} --- play a crucial role for the nucleon spin structure. 
Both functions describe the strength of spin-orbit interactions that are similar to spin-orbit interactions in atomic systems like hydrogen~\cite{Lorce:2011kd,Lorce:2014mxa}.
In particular, there is a direct relation between $F_{1,4}$ and the orbital angular momentum (OAM) of partons inside a longitudinally polarized nucleon~\cite{Lorce:2011kd,Hatta:2011ku,Ji:2012sj}.
It is remarkable that the same relation between $F_{1,4}$ and the quark OAM holds for both commonly used OAM definitions --- the (canonical) one by Jaffe and Manohar $(L_{\small \textrm{JM}})$~\cite{Jaffe:1989jz}, and the one by Ji $(L_{\small \textrm{Ji}})$~\cite{Ji:1996ek}.
This representation of OAM also allows for an intuitive interpretation of the difference between $L_{\small \textrm{JM}}$ and $L_{\small \textrm{Ji}}$~\cite{Burkardt:2012sd}.
Moreover, it gives access to the so far elusive $L_{\small \textrm{JM}}$ in quantum chromodynamics (QCD) on the lattice~\cite{Hatta:2011ku,Engelhardt:2017miy,Rajan:2016tlg}.

While a number of model calculations of GTMDs is available by
now~\cite{Meissner:2008ay, Meissner:2009ww, Lorce:2011dv, Lorce:2011kd, Lorce:2011ni, Kanazawa:2014nha, Mukherjee:2014nya, Hagiwara:2014iya, Lorce:2015sqe, Chakrabarti:2016yuw, Echevarria:2016mrc, Hagiwara:2016kam, Gutsche:2016gcd, Zhou:2016rnt, Courtoy:2016des, More:2017zqq}, for many years it was unknown how GTMDs can be measured. 
Only recently it was shown that GTMDs of gluons can, in principle, be accessed via diffractive di-jet production in deep-inelastic lepton-nucleon and lepton-nucleus scattering~\cite{Hatta:2016dxp,Hatta:2016aoc,Ji:2016jgn}, as
well as virtual photon-nucleus quasi-elastic scattering~\cite{Zhou:2016rnt}. 
Some numerical studies of gluon GTMDs at small $x$, based on a saturation model, were performed in Refs.~\cite{Hagiwara:2016kam,Zhou:2016rnt}.
Not long ago it was also pointed out that gluon GTMDs can be studied in proton-nucleus collisions~\cite{Hagiwara:2017ofm}. 
With the exception of~\cite{Ji:2016jgn}, the papers on observables for GTMDs deal with the small-$x$ region of parton saturation.

In this work we identify, for the first time, a physical process which gives access to quark GTMDs.
Specifically, we show how GTMDs enter the exclusive pion-nucleon double Drell-Yan process,
$\pi N \to (\ell_1^- \ell_1^+) (\ell_2^- \ell_2^+) N'$, where one detects two di-lepton pairs plus a nucleon.
To this end, we perform a leading-order (LO) analysis in perturbative QCD.
Our main focus is on the GTMDs $F_{1,4}$ and $G_{1,1}$, which can be measured through suitable polarization observables.
We also argue that other quark GTMDs could be systematically studied in the same process.

\section{Generalized TMDs} 
\label{s:gtmd}
Let us briefly recall the definition of quark GTMDs for a nucleon~\cite{Meissner:2008ay,Meissner:2009ww}, to the extent it is necessary for the present work.
GTMDs parameterize the off-forward transverse momentum dependent quark-quark correlator
\begin{equation} \label{e:gtmd_corr}
W_{\lambda,\lambda'}^{q \, [\Gamma]} (P,\Delta,x,\vec{k}_\perp) = 
\int \frac{dz^- \, d^2\vec{z}_\perp}{2 (2\pi)^3} \, e^{i k \cdot z} \, 
\langle p', \lambda' | \, \bar{q}(- \tfrac{z}{2}) \, \Gamma \, {\cal W}(- \tfrac{z}{2}, \tfrac{z}{2}) \, q(\tfrac{z}{2}) \, | p, \lambda \rangle \Big|_{z^+ = 0} \,,
\end{equation}
where $q$ indicates the quark flavor and $\Gamma$ a generic gamma matrix.
The 4-momenta and the helicities of the incoming (outgoing) nucleon are denoted by $p (p')$ and $\lambda (\lambda')$, respectively.
We also use the definitions $P = (p + p')/2$ and $\Delta = p' - p$.
The two quark fields of the operator in~(\ref{e:gtmd_corr}) are separated along the light-cone minus direction $z^-$ and the transverse direction $\vec{z}_\perp$.
(We define the light-cone components of a generic 4-vector $a = (a^0, a^1, a^2, a^3)$ through $a^\pm = (a^0  \pm a^3) / \sqrt{2}$ and $\vec{a}_\perp = (a^1, a^2)$.) 
The Wilson line $\cal W$ makes the bi-local operator color gauge invariant.
The average longitudinal and transverse quark momenta are given by $x$ and $\vec{k}_\perp$, respectively.
We also point out that, strictly speaking, some modification of the definition in~(\ref{e:gtmd_corr}) is needed in order to avoid the infamous light-cone singularities.
More information on this point, which is irrelevant for the main purpose of the present work, can be found in~\cite{Echevarria:2016mrc} and references therein.

Here we need the parametrization of~(\ref{e:gtmd_corr}) in terms of GTMDs for $\Gamma = \gamma^+, \gamma^+ \gamma_5$.
In the notation of~\cite{Meissner:2009ww} they read
\begin{eqnarray} \label{e:gammap}
W_{\lambda,\lambda'}^{q \, [\gamma^+]} & = & \frac{1}{2M} \, \bar{u}(p',\lambda') \bigg[ 
F_{1,1}^q + \frac{i  \sigma^{i+}  k_\perp^i}{P^+} \, F_{1,2}^q + \frac{i \, \sigma^{i+} \Delta_\perp^i}{P^+} \, F_{1,3}^q 
+ \frac{i \sigma^{ij} k_{\perp}^i \Delta_{\perp}^j}{M^2} \, F_{1,4}^q  \bigg] u(p,\lambda)
\nonumber \\
& = & \frac{1}{M \sqrt{1 - \xi^2}} \bigg\{ 
\bigg[ M \delta_{\lambda,\lambda'} - \frac{1}{2} \Big( \lambda \Delta_\perp^1 + i \Delta_\perp^2 \Big) \delta_{\lambda,-\lambda'} \bigg] F_{1,1}^q
+ (1 - \xi^2) \Big( \lambda k_\perp^1 + i k_\perp^2 \Big) \delta_{\lambda,-\lambda'} \, F_{1,2}^q
\nonumber \\
& & \hspace{0.8cm} 
+ \; (1 - \xi^2) \Big( \lambda \Delta_\perp^1 + i \Delta_\perp^2 \Big) \delta_{\lambda,-\lambda'} \, F_{1,3}^q
+ \frac{i \varepsilon_\perp^{ij} k_{\perp}^i \Delta_{\perp}^j}{M^2} \bigg[ \lambda M \delta_{\lambda,\lambda'} - \frac{\xi}{2} \Big( \Delta_\perp^1 + i \lambda \Delta_\perp^2 \Big) \delta_{\lambda,-\lambda'} \bigg] F_{1,4}^q \bigg\} \,,
\\[0.2cm]  \label{e:gammap5}
W_{\lambda,\lambda'}^{q \, [\gamma^+ \gamma_5]} & = & \frac{1}{2M} \, \bar{u}(p',\lambda') \bigg[ 
- \frac{i \varepsilon_\perp^{ij} k_{\perp}^i \Delta_{\perp}^j}{M^2} \, G_{1,1}^q
+ \frac{i  \sigma^{i+}  \gamma_5 k_\perp^i}{P^+} \, G_{1,2}^q + \frac{i  \sigma^{i+}  \gamma_5 \Delta_\perp^i}{P^+} \, G_{1,3}^q
+ i \sigma^{+-} \gamma_5 \, G_{1,4}^q  \bigg] u(p,\lambda)
\nonumber \\
& = & \frac{1}{M \sqrt{1 - \xi^2}} \bigg\{
- \frac{i \varepsilon_\perp^{ij} k_{\perp}^i \Delta_{\perp}^j}{M^2} \bigg[ M \delta_{\lambda,\lambda'} - \frac{1}{2} \Big( \lambda \Delta_\perp^1 + i \Delta_\perp^2 \Big) \delta_{\lambda,-\lambda'} \bigg] G_{1,1}^q
+ (1 - \xi^2) \Big( k_\perp^1 + i \lambda k_\perp^2 \Big) \delta_{\lambda,-\lambda'} \, G_{1,2}^q
\nonumber \\
& & \hspace{0.8cm} 
+ \; (1 - \xi^2) \Big( \Delta_\perp^1 + i \lambda \Delta_\perp^2 \Big) \delta_{\lambda,-\lambda'} \, G_{1,3}^q
+ \bigg[ \lambda M \delta_{\lambda,\lambda'} - \frac{\xi}{2} \Big( \Delta_\perp^1 + i \lambda \Delta_\perp^2 \Big) \delta_{\lambda,-\lambda'} \bigg] G_{1,4}^q \bigg\} \,.
\end{eqnarray}
In order to evaluate the matrix elements in the first lines of Eqs.~(\ref{e:gammap}),~(\ref{e:gammap5}) we considered $u(p,\lambda)$ and $u(p',\lambda')$ as light-cone helicity spinors~\cite{Soper:1972xc,Diehl:2003ny}.
Note that $M$ is the nucleon mass, and $\xi = (p^+ - p'^+) / (p^+ + p'^+) = - \Delta^+ / (2 P^+)$ characterizes the longitudinal momentum transfer to the nucleon.
We also use $\sigma^{\mu\nu} = i [\gamma^\mu, \gamma^\nu] / 2$, and $\varepsilon_\perp^{ij} = \varepsilon^{-+ij}$ with $\varepsilon^{0123} = 1$.
The kinematical arguments on the l.h.s.~of (\ref{e:gammap}),~(\ref{e:gammap5}) are suppressed.
For a generic GTMD one has $X = X(x, \xi, \vec{k}_\perp, \vec{\Delta}_\perp)$, where the dependence on $\vec{k}_\perp$ and $\vec{\Delta}_\perp$ is through the scalar products which can be formed by these vectors.
We also recall that, in general, GTMDs are complex-valued functions~\cite{Meissner:2008ay,Meissner:2009ww}.
For the reasons given above our main focus will be on the GTMDs $F_{1,4}$ and $G_{1,1}$.
The real part of the GTMDs $F_{1,1}$ and $G_{1,4}$ has a close connection to the distribution of unpolarized quarks in an unpolarized nucleon and the distribution of longitudinally polarized quarks in a longitudinally polarized nucleon, respectively~\cite{Meissner:2009ww,Lorce:2011kd}.
Since these distributions are large we will also consider observables which are sensitive to their interference with $F_{1,4}$ and $G_{1,1}$. 
Below we will concentrate on the helicity-conserving terms in~(\ref{e:gammap}),~(\ref{e:gammap5}) that are proportional to $\delta_{\lambda,\lambda'}$.

The cross section for the double Drell-Yan process is also sensitive to the matrix element
\begin{equation} \label{e:pion_corr}
\Phi^q(x,\vec{k}_\perp^{\,2}) = \int \frac{dz^+ \, d^2\vec{z}_\perp}{2 (2\pi)^3} \, e^{i (k - p/2) \cdot z} \, 
\langle 0| \, \bar{q}(- \tfrac{z}{2}) \, \gamma^- \gamma_5 \, {\cal W}(- \tfrac{z}{2}, \tfrac{z}{2}) \, q(\tfrac{z}{2}) \, | \pi(p) \rangle \Big|_{z^- = 0} \,.
\end{equation}
Modulo pre-factors, $\Phi^q(x,\vec{k}_\perp^{\,2})$ is the light-cone wave function of the pion~\cite{Diehl:2003ny,Brodsky:1997de}.
The double Drell-Yan process implies in both Eq.~(\ref{e:gtmd_corr}) and Eq.~(\ref{e:pion_corr}) a staple-like past-pointing Wilson line~\cite{BMZ:prep}, identical to the one that appears in TMD factorization of the ordinary Drell-Yan process~\cite{Collins:2002kn,Ji:2002aa,Belitsky:2002sm}.

\section{Double Drell-Yan process and polarization observables} 
\label{s:doubleDY}
\begin{figure}[t]
\begin{center}
\includegraphics[width=14.0cm]{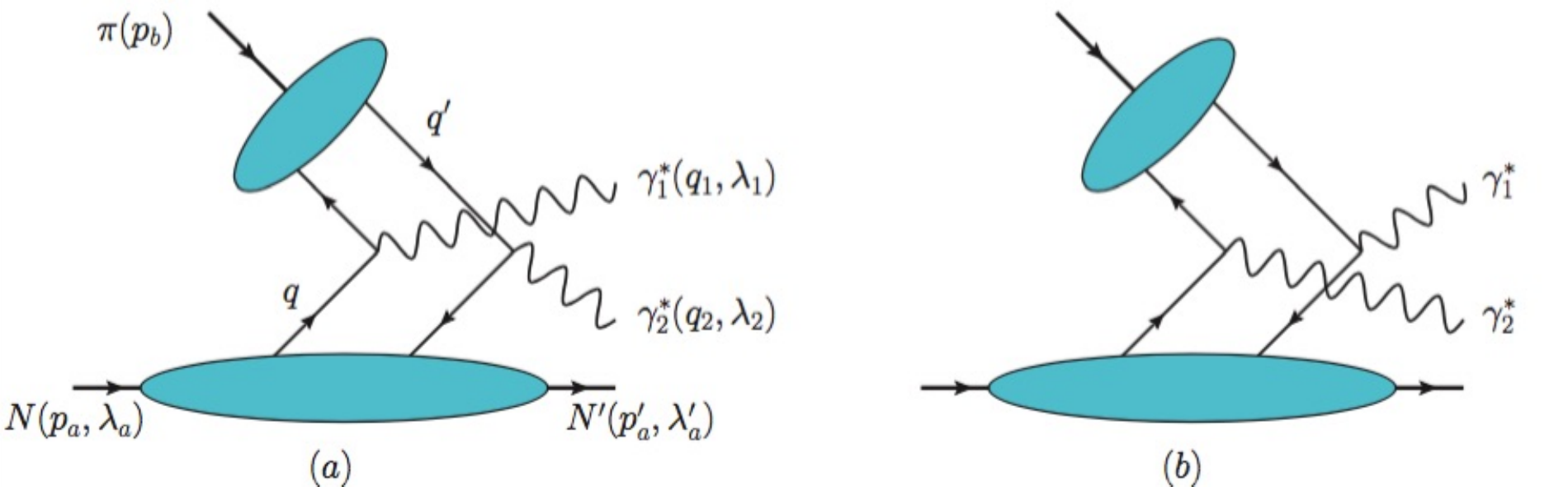}
\end{center}
\caption{LO diagrams for the exclusive double Drell-Yan process $\pi \, N \to \gamma_1^{\ast} \gamma_2^{\ast} N'$.}
\label{f:piN_doubleDY}
\end{figure}

To calculate observables we consider the production of two virtual photons rather than two di-lepton pairs.
Specifically, we study the process
\begin{equation}
\pi(p_b) + N(p_a,\lambda_a) \to \gamma_1^{\ast}(q_1,\lambda_1) + \gamma_2^{\ast}(q_2,\lambda_2)+N'(p_a',\lambda_a') \,.
\end{equation}
From here on the variables of the incoming and outgoing nucleon carry an index $a$ compared to above.
We concentrate on large $s = (p_a + p_b)^2 \approx 2 p_a^+ p_b^-$, large photon virtualities $q_1^2$, $q_2^2$, and small transverse photon momenta, $|\vec{q}_{i\perp}^{\; 2}| \ll q_i^2$.
In this region one can use TMD-type factorization.
The longitudinal momentum transfer to the nucleon can be written as $\xi_a = (q_1^+ + q_2^+)/(2P_a^+)$.
The LO diagrams for this process are shown in Fig.$\,$\ref{f:piN_doubleDY}.
The scattering amplitude depends on the helicities of the nucleons and photons,
\begin{equation} \label{e:amp_1}
{\cal T}_{\lambda_a, \lambda_a'}^{\lambda_1, \lambda_2} = 
{\cal T}_{\lambda_a, \lambda_a'}^{\mu \nu} \, \varepsilon_\mu^\ast(\lambda_1) \, \varepsilon_\nu^\ast(\lambda_2) \,,
\end{equation}
where $\varepsilon^{\mu}(\lambda_1)$  and $\varepsilon^{\mu}(\lambda_2)$ are the photon polarization vectors.
One finds
\begin{eqnarray} \label{e:amp_2}
{\cal T}_{\lambda_a, \lambda_a'}^{\mu\nu} & = &
 i \sum_{q,q' }e_q e_q' e^2 \frac{(2 \pi)^4}{N_c} \int d^2 \vec{k}_{a\perp} \int d^2 \vec{k}_{b\perp} \delta^{(2)} \bigg( \frac{\Delta \vec{q}_\perp}{2} - \vec{k}_{a\perp} - \vec{k}_{b\perp} \bigg) 
 \Phi_\pi^{q' q}(x_b,\vec{k}_{b\perp}^{\,2})
\nonumber \\
& & \bigg[ - i \varepsilon_\perp^{\mu\nu} \Big( W_{\lambda_a, \lambda_a'}^{qq' \, [\gamma^+]}(x_a, \vec{k}_{a\perp}) 
- W_{\lambda_a, \lambda_a'}^{qq' \, [\gamma^+]}(- x_a, - \vec{k}_{a\perp}) \Big)
\nonumber \\
& &\hspace{0.3cm} - \, g_\perp^{\mu\nu} \Big( W_{\lambda_a, \lambda_a'}^{qq' \, [\gamma^+ \gamma_5]}(x_a, \vec{k}_{a\perp}) 
+ W_{\lambda_a, \lambda_a'}^{qq' \, [\gamma^+ \gamma_5]}(- x_a, - \vec{k}_{a\perp}) \Big) 
\bigg] \,,
\end{eqnarray}
where $e_q$ and $e_q'$ are the quark charges in units of the elementary charge $e$, and $N_c$ is the number of quark colors.
The expression in~(\ref{e:amp_2}) describes the double Drell-Yan process for all possible pion and nucleon charge states.
Note that $\Phi_\pi^{q'q}$ is defined as in~(\ref{e:pion_corr}), but with the operator $\bar{q}' \gamma^- \gamma_5 \, q$.
Isospin symmetry provides $\Phi_{\pi^+}^{du} = \Phi_{\pi^-}^{ud} = \sqrt{2} \, \Phi_{\pi^0}^{uu} = - \sqrt{2} \, \Phi_{\pi^0}^{dd}$.
Likewise, $W^{qq' [\Gamma]}$ is given by~(\ref{e:gtmd_corr}) with the operator $\bar{q} \, \Gamma \, q'$. 
With this notation one can also describe transitions between different nucleons.
Like in the case of transition GPDs, for the GTMDs one has $X_{p \to n}^{du} = X_{n \to p}^{ud}  = X_{p}^{u} - X_{p}^{d}$~\cite{Mankiewicz:1997aa}.
In Eq.~(\ref{e:amp_2}) we use the vector $\Delta\vec{q}_\perp = \vec{q}_{1\perp} - \vec{q}_{2\perp}$. 
The transverse momenta of the photons can be expressed by $\Delta\vec{q}_\perp$ and the transverse momentum transfer to the nucleon $\vec{\Delta}_{a\perp} = -(\vec{q}_{1\perp} + \vec{q}_{2\perp})$.
While the amplitude contains an integration upon the transverse momenta of the quarks, their longitudinal momenta are fixed according to
$x_a = (q_1^+ - q_2^+) /(2 P_a^+), \; x_b = 1 - q_1^- / p_b^-  = q_2^- / p_b^-$.
The value for $x_a$ implies the so-called ERBL region~\cite{Efremov:1979qk,Lepage:1979zb}, characterized by $- \xi_a \le x_a \le \xi_a$, in which the GTMD matrix element describes the emission of a quark-antiquark pair from the nucleon.
The amplitude, {\it a priori}, depends on both the  $F_{1,i}$ and the $G_{1,i}$ $(i = 1, \ldots, 4)$.
From~(\ref{e:amp_2}) one readily sees that the dominant contribution to the amplitude is for transversely polarized photons.
In this context note that $g_{\perp}^{\mu\nu} = g^{\mu\nu} - n_a^\mu n_b^\nu - n_a^\nu n_b^\mu$, with the light-like vectors $n_a = (1,0,0,-1)/\sqrt{2}, \; n_b = (1,0,0,1)/\sqrt{2}$.

The relation between the scattering amplitude in~(\ref{e:amp_1}) and the cross section in the center-of-mass frame reads
\begin{equation}
d \sigma_{\lambda_a, \lambda_a'}^{\lambda_1, \lambda_2} = \frac{\pi}{2 s^{3/2}} \frac{1 + \xi_a}{1 - \xi_a}|{\cal T}_{\lambda_a, \lambda_a'}^{\lambda_1, \lambda_2}|^2 \delta(p_a'^0 + q_1^0 + q_2^0 - \sqrt{s}) \frac{d^4 q_1}{(2\pi)^4} \frac{d^4 q_2}{(2\pi)^4} \,,
\end{equation}
where we have already integrated over the phase space of the outgoing nucleon.
Below we consider the unpolarized cross section, single-spin asymmetries (SSAs), and double-spin asymmetries (DSAs).
It is convenient to introduce 
\begin{eqnarray} \label{e:tau_uu}
\tau_{UU} & = & \frac{1}{2} \sum_{\lambda,\lambda'} |{\cal T}_{\lambda,\lambda'}|^2 \,,
\\ \label{e:tau_lu}
\tau_{LU} & = & \frac{1}{2} \sum_{\lambda'} \Big( |{\cal T}_{+,\lambda'}|^2 - |{\cal T}_{-,\lambda'}|^2 \Big) \,,
\\ \label{e:tau_ll}
\tau_{LL} & = & \frac{1}{2} \Big( \big(|{\cal T}_{+,+}|^2 - |{\cal T}_{+,-}|^2 \big)  - \big(|{\cal T}_{-,+}|^2 - |{\cal T}_{-,-}|^2 \big)\Big) \,,
\end{eqnarray}
where summation over the photon polarizations is implied.
Obviously, $\tau_{LU}$ determines the numerator of the longitudinal target SSA, whereas $\tau_{LL}$ describes the longitudinal DSA with polarization of both the target and the recoil nucleon.
Spin asymmetries for transverse polarization in the $x$-direction or $y$-direction are defined accordingly.

In order to get direct access to $F_{1,4}$, that is, without interference with other GTMDs, one has to consider a linear combination of (polarization) observables, 
\begin{eqnarray} \label{e:pol_obs_1}
\frac{1}{4} \big( \tau_{UU} + \tau_{LL} - \tau_{XX} - \tau_{YY} \big) & = &
\frac{2}{M^4} \big( \varepsilon_\perp^{ij} \Delta q_\perp^i \Delta_{a\perp}^j \big)^2 \,
C^{(+)} \Big[ \vec{\beta}_\perp \cdot \vec{k}_{a\perp} \, F_{1,4} \, \Phi_\pi \Big]
C^{(+)} \Big[ \vec{\beta}_\perp \cdot \vec{k}_{a\perp} \, F_{1,4}^\ast \, \Phi_\pi^\ast \Big]
\nonumber \\[0.2cm]
& & + \; 2 \, C^{(+)} \Big[ G_{1,4} \, \Phi_\pi \Big]
C^{(+)} \Big[ G_{1,4}^\ast \, \Phi_\pi^\ast \Big] \,.
\end{eqnarray}
In Eq.~(\ref{e:pol_obs_1}) we use the shorthand notation 
\begin{eqnarray}
C^{(\pm)} \Big[ w(\vec{k}_{a\perp},\vec{k}_{b\perp}) \, X \, \Phi_\pi \Big] & = & 
\frac{e^2}{\sqrt{1 - \xi_a^2} N_c} \sum_{q,q'} e_q e_q' \int d^2 \vec{k}_{a\perp} \int d^2 \vec{k}_{b\perp} \,
\delta^{(2)}\bigg( \frac{\Delta \vec{q}_\perp}{2} - \vec{k}_{a\perp} - \vec{k}_{b\perp} \bigg) w(\vec{k}_{a\perp},\vec{k}_{b\perp}) 
\nonumber \\[0.2cm]
& & \times \; \Big[ X^{qq'}(x_a,\vec{k}_{a\perp}) \pm X^{qq'}(-x_a,-\vec{k}_{a\perp}) \Big]
\Phi_\pi^{q' q}(x_b,\vec{k}_{b\perp}^{\,2}) \,,
\end{eqnarray}
with $w(\vec{k}_{a\perp},\vec{k}_{b\perp})$ a generic weight function.
The vector $\vec{\beta}_\perp$ in~(\ref{e:pol_obs_1}) reads
\begin{equation}
\vec{\beta}_\perp = \frac{ \vec{\Delta}_{a\perp}^2 \, \Delta \vec{q}_\perp - (\vec{\Delta}_{a\perp} \cdot \Delta \vec{q}_\perp) \, \vec{\Delta}_{a\perp}}{\vec{\Delta}_{a\perp}^2 \, \Delta \vec{q}_\perp^{\; 2} - (\vec{\Delta}_{a\perp} \cdot \Delta \vec{q}_\perp)^2} \,.
\end{equation}  
We repeat that in order to obtain Eq.~(\ref{e:pol_obs_1}) the photon polarizations have been summed over.
While in that case there is no interference between $F_{1,4}$ and other GTMDs, one still has a second term which is given by $G_{1,4}$.
As already mentioned, $G_{1,4}$ is presumably large, and therefore it may actually be difficult to address $F_{1,4}$ through this observable, unless one has a reliable estimate of $G_{1,4}$.
However, one can separate the two contributions in~(\ref{e:pol_obs_1}) by not summing over the photon polarizations.
For instance, if one projects on appropriate linear polarizations of the photons, the contributions of either $F_{1,4}$ or $G_{1,4}$ can be switched off~\cite{BMZ:prep}.
This result, which holds irrespective of the polarization states of the nucleons, follows from the expression in~(\ref{e:amp_2}).
To address $G_{1,1}$ one can study $\frac{1}{4} \big( \tau_{UU} + \tau_{LL} + \tau_{XX} + \tau_{YY} \big)$.
The result for this linear combination is identical to (\ref{e:pol_obs_1}), but with the replacements $F_{1,4} \to G_{1,1}$ and $G_{1,4} \to F_{1,1}$.
Again, the contributions from $G_{1,1}$ and $F_{1,1}$ can be separated by measuring suitable photon polarizations.

Apart from the fact that a considerable number of different polarization measurements is required, the observable in~(\ref{e:pol_obs_1}), as well as the corresponding observable for $G_{1,1}$, may have a drawback: 
In these linear combinations one has cancellations of potentially large terms~\cite{BMZ:prep}.
It may therefore be beneficial to also explore interference between $F_{1,4}$ (or $G_{1,1}$) and other GTMDs.
Such an interference shows up in the following linear combination of longitudinal SSAs:
\begin{eqnarray} \label{e:pol_obs_2}
\frac{1}{2} \big( \tau_{LU} + \tau_{UL} \big) & = &  \frac{1}{2} \big( |{\cal T}_{+,+}|^2 - |{\cal T}_{-,-}|^2  \big)
= \frac{4}{M^2} \, \varepsilon_\perp^{ij} \Delta q_\perp^i \Delta_{a\perp}^j  \,
\textrm{Im} \, \bigg\{ C^{(-)} \Big[  F_{1,1} \, \Phi_\pi \Big]
C^{(+)} \Big[ \vec{\beta}_\perp \cdot \vec{k}_{a\perp} \, F_{1,4}^\ast \, \Phi_\pi^\ast \Big]
\nonumber \\
&& \hspace{3.7cm} -  \; C^{(+)} \Big[ G_{1,4} \, \Phi_\pi \Big]
C^{(-)} \Big[ \vec{\beta}_\perp \cdot \vec{k}_{a\perp} \, G_{1,1}^\ast \, \Phi_\pi^\ast \Big]
\bigg \} \,.
\end{eqnarray}
We point out that the expressions for $\tau_{LU}$ or $\tau_{UL}$ alone are more complicated as they contain additional GTMDs~\cite{BMZ:prep}.
More polarization observables exist which involve interference between $F_{1,4}$ (or $G_{1,1}$) and other GTMDs, but the observable in~(\ref{e:pol_obs_2}) gives the simplest expression~\cite{BMZ:prep}.
Note that on the r.h.s.~of~(\ref{e:pol_obs_2}) the imaginary part of products of GTMDs appears.
According to current knowledge the GTMDs most relevant for the spin structure of the nucleon are $\textrm{Re} \, F_{1,4}$ and $\textrm{Re} \, G_{1,1}$.
Though these functions contribute to (\ref{e:pol_obs_2}) they interfere with $\textrm{Im} \, F_{1,1}$ and $\textrm{Im} \, G_{1,4}$, respectively.
At present, there exists no information on the latter functions, and they may in fact be small.
This issue can be overcome by considering the observable $\frac{1}{2} \big( \tau_{XY} - \tau_{YX} \big)$, whose result agrees with the r.h.s.~of~(\ref{e:pol_obs_2}) but with $\textrm{Re} \, \{\ldots \}$ instead of $\textrm{Im} \, \{\ldots \}$.

\section{Conclusions} 
\label{s:concl}
We have shown that GTMDs for quarks can be studied through the exclusive double Drell-Yan process.
Specifically, to leading order in perturbative QCD, this process in sensitive to GTMDs in the ERBL region. 
The main focus was on the GTMDs $F_{1,4}$ and $G_{1,1}$ which recently attracted much attention because of their relation to the spin structure of the nucleon.
The double Drell-Yan process leads to a staple-like Wilson line for the operator definition of the GTMDs~\cite{BMZ:prep}, providing the connnection to the canonical OAM $(L_{\small \textrm{JM}})$~\cite{Hatta:2011ku}. 
We have proposed several polarization observables which can give access to these GTMDs, either directly or through interference with other GTMDs. 
In a similar manner, all leading-twist GTMDs could be explored via suitable polarization observables~\cite{BMZ:prep}.

Several extensions of our work can be envisioned.
An attempt should be made to numerically estimate both the unpolarized cross section and the various spin asymmetries to find out if the reaction $\pi N \to (\ell_1^- \ell_1^+) (\ell_2^- \ell_2^+) N'$ is measurable at existing facilities.
Also, one can perform a similar analysis for nucleon-nucleon collisions~\cite{BMZ:prep}.
Production of heavy gauge bosons instead of photons may be considered as well.
Moreover, hadronic final states typically give rise to higher count rates.
One such example is the process $pp \to \eta_c \eta_c pp$, which can basically be treated along the lines discussed above, though gluon GTMDs enter the leading-order analysis~\cite{BMZ:prep}.
We finally point out that the type of reactions discussed here could also provide constraints on GPDs in the ERBL region, where experimental information is still sparse --- see Refs.~\cite{Berger:2001zn,Goloskokov:2015zsa,Sawada:2016mao} for related work on GPDs.
\\
\begin{acknowledgments}
This work has been supported by the National Science Foundation under Contract No.~PHY-1516088 (A.M.), the National Science Foundation of China under Grant No.~11675093, and by the Thousand Talents Plan for Young Professionals (J.Z.).
The work of A.M.~was supported by the U.S. Department of Energy, Office of Science, Office of Nuclear Physics, within the framework of the TMD Topical Collaboration.
\end{acknowledgments}


\end{document}